\documentclass[12pt]{article}
\usepackage{amsmath,amssymb}
\usepackage{latexsym,graphicx,epsfig}
\date{}
\newcommand{\fr}[2]{\frac{{\displaystyle #1}}{{\displaystyle #2}}}

\def\epe{\mbox{$e^+e^-\,$}}
\def\ggam{\mbox{$\gamma\gamma\,$}}
\def\egam{\mbox{$e\gamma \,$}}

\def\egeh{\mbox{$e\gamma\to eh\,$}}
    \newcommand{\fn}[1]{\footnote{{\normalsize #1}}}

\def\noCP{$\rlap{CP}\,\diagup$\ \ }

    \newenvironment{Itemize}{\begin{list}{$\bullet$}%
    {\setlength{\topsep}{0.2mm}\setlength{\partopsep}{0.2mm}%
    \setlength{\itemsep}{0.2mm}\setlength{\parsep}{0.2mm}}}%
    {\end{list}}
    \newcounter{enumct}

\newsavebox{\fmbox}
    \newenvironment{fmpage}[1]
    {\begin{lrbox}{\fmbox}\begin{minipage}{#1}}
    {\end{minipage}\end{lrbox}\fbox{\usebox{\fmbox}}}
    
    \def\NIM{{\em Nucl. Instrum. Methods\ }}

\title{Two Higgs Doublet Model, Standard Model--like scenario and
resolving SM from 2HDM at Photon Colliders\\ {\em\normalsize
Talk given at XVI Workshop on High Energy Physics and Quantum
Field Theory, Moscow, September 2001 }}

 \author{I.F. Ginzburg\thanks{e-mail: ginzburg@math.nsc.ru},  M.V. Vychugin \\
Sobolev Institute of Mathematics and Novosibirsk State
University,\\ Novosibirsk, Russia}

\begin{document}

 \maketitle

\begin{abstract}

1. Discussing 2HDM and Higgs potential, we find the parameters
range giving {\em naturally} {\bf weak} effects of CP violation
and Flavor Changing Neutral Currents. Within this approach, the
widely discussed decoupling limit with heavy $H^\pm$, $H$ and $A$
corresponds to an unnatural set of 2HDM parameters.

2. We define the SM--like scenario as that in which after  LHC and
\epe\ LC experiments no visible deviations from the minimal SM
will be found. We find all the parameter ranges of 2HDM compatible
with this scenario.

3. We show that the study of Higgs boson production at Photon
Colliders (processes $\ggam\to h$ and \egeh) can help to
distinguish minimal SM from 2HDM with natural set of parameters.
\end{abstract}

\section{  Two Higgs Doublet Model }

The different variants of Higgs type spontaneous electroweak
symmetry breaking (EWSB) in the Standard Model are described by
lagrangian\fn{{\it\normalsize This part is based on the paper
\cite{GKO2}.}}
 \begin{equation}\begin{array}{l}
{ \cal L}={ \cal L}^{SM}_{gf} +{ \cal L}_H +{ \cal L}_Y+  V
\,;\\[2mm]
 { \cal L}^{SM}_{gf}\; \mbox{ --  SM interaction of
gauge bosons and fermions}\,,\\[2mm]
 {\cal L}_H
 =\sum \limits_a(D_{\mu} \phi_a )^{ \dagger}(D_{\mu} \phi_a)
\;\mbox{ -- scalar kinetic term} \,, \\[2mm]
 {\cal L}_Y\; \mbox{ -- Yukawa interaction of
fermions to scalars}\,,\\[2mm]
 V\;\mbox{ -- Higgs potential}\,.
 \end{array}\label{lagrbas}
 \end{equation}
In the Minimal Standard Model (SM) the single Higgs isodoublet
($a=1$) with hypercharge $Y=1$ is used. The simplest extension of
SM is the Two Higgs Doublet Model (2HDM) with 2 complex scalar
doublets ($a=1,2$). To keep for the quantity $\rho =M_W^2/ (M_Z^2
\cos^2\theta_W)$ its SM value $\rho= 1$ at tree level, both Higgs
fields should be naturally weak isodoublets ($T=1/2$) with
hypercharges $Y= \pm1$. We use $Y=+1$ for both (the other choices,
e.g. $Y_1=1,\;Y_2=-1$ -- as in MSSM -- keep our results up to
redefinitions).

The Higgs sector, different from that in the minimal SM, generally
gives CP violation (\noCP\hspace{-2mm}) and Flavor Changing
Neutral Currents (FCNC) at EWSB. In the 2HDM the
$(\phi_1,\,\phi_2)$ mixing plays a crucial role in these
violations. Let us summarize main points.\vspace{2mm}

\noindent
 \begin{fmpage}{0.99\textwidth}{
$\Box$ The Higgs potential generates CP violation only if
$(\phi_1,\,\phi_2)$ mixing exists, and corresponding coefficients
in potential are complex. We discuss this point in details below.

$\Box$ $(\phi_1,\,\phi_2)$ mixing can also be originated from
the Yukawa type interaction -- if any fermions couple to both
scalar fields, the one--loop polarization operator generates
the $(\phi_1,\,\phi_2)$ mixing. Such Yukawa interaction naturally
contains terms which are off--diagonal in family index
giving potentially large FCNC effects.}
\end{fmpage}
\vspace{2mm}

It is well known that both \noCP and FCNC effects are weak.
Therefore, the {\em natural} construction of 2HDM should start
with the lagrangian having an additional symmetry forbidding the
\noCP and FCNC effects, or ($\phi_1,\phi_2$) mixing. This is
{\bf\boldmath $Z_2$ symmetry} at
 \begin{equation}
\phi_1 \leftrightarrow-\phi_1,\; \phi_2
\leftrightarrow\phi_2\;\mbox{ and vice versa.} \label{Z2sym}
 \end{equation}
This symmetry can be weakly broken to allow weak \noCP and FCNC
effects. Therefore {\bf\boldmath the terms in the lagrangian
giving ($\phi_1,\phi_2$) mixing have to be absent or small.}

$\bullet$ Below we use ratios of actual coupling constants of each
neutral Higgs scalar $\phi$ to   particle $i$ to corresponding
values for the Higgs boson in the SM,
 \begin{subequations}
\label{Eq:chi-def}
 \begin{equation}
\chi_i^\phi= \fr{g_i^\phi}{g_i^{\rm SM}}\Rightarrow
\chi_{fS}^{\phi}+i\gamma^5 \chi_{fA}^{\phi} \,.\label{chidef}
 \end{equation}
The latter form arises for fermions if Higgs--like boson has no
definite CP parity, being a mixture of scalar and pseudoscalar.

If fermion mass $M_f$ is much lower than the mass of considered
Higgs boson $M_i$, the partial width of corresponding decay is
given practically by the quantity $|\chi_f^2|$,
 \begin{equation}
\Gamma(h_i\to f\bar{f})=\Gamma^{SM}(h\to f\bar{f})
 \left[|\chi_f^i|^2+|\chi_{fA}^i|^2
{\cal O}\left(\fr{M_f^2}{M_i^2}\right)\right]\,.
 \end{equation}
 \end{subequations}

\subsection{Higgs potential}

The most general Higgs potential in a renormalizable theory can be
written as
 \begin{equation}\begin{array}{c}
V=\fr{\lambda_1}{2}(\phi_1^\dagger\phi_1)^2
+\fr{\lambda_2}{2}(\phi_2^\dagger\phi_2)^2+
\lambda_3(\phi_1^\dagger\phi_1) (\phi_2^\dagger\phi_2)\\[2mm]
+\lambda_4(\phi_1^\dagger\phi_2) (\phi_2^\dagger\phi_1)
 +\fr{1}{2}\left[\lambda_5(\phi_1^\dagger\phi_2)^2
+h.c.\right]+\Delta V_m^4+{\cal M}(\phi_i)+V_0\,;\\[2mm]
 \Delta V_m^4
=\left\{\left[\lambda_6(\phi_1^\dagger\phi_1)+\lambda_7
(\phi_2^\dagger\phi_2)\right](\phi_1^\dagger\phi_2)
+h.c.\right\},\\[3mm]
 {\cal M}(\phi_i)=-\fr{1}{2}\left\{m_{11}^2(\phi_1^\dagger\phi_1)+
\left[m_{12}^2 (\phi_1^\dagger\phi_2) +h.c.\right]+
m_{22}^2(\phi_2^\dagger\phi_2)\right\},\quad V_0=const\,.
\end{array}\label{baspot}
 \end{equation}
Here $\lambda_{1-4}$, $m_{11}^2$ and $m_{22}^2$ are real, while
parameters $\lambda_{5-7}$ and $m_{12}$ are generally complex. The
constant $V_0$ is added to make vacuum energy equal to  zero after
EWSB.

To eliminate $(\phi_1,\,\phi_2)$ mixing, one should set $m_{12}=0$
and  $\Delta V_m^4=0$. The term with $m_{12}$ describes {\em soft
violation} of $Z_2$ symmetry. In our discussion limited to the
tree level, additional contribution $\Delta V_m^4$ introduces no
new phenomenology, it just complicates  the equations. At the loop
level the term $\Delta V_m^4$ causes certain difficulties in
description of the model. By these reasons we omit this term in
the forthcoming discussion (as  many authors do "for the sake of
simplicity").

$\Box$ The minimum of the potential defines vacuum expectation
values $\langle\phi_i\rangle$ (v.e.v.'s) of the fields $\phi_i$
via equations
\begin{subequations}
\label{vev}
\begin{equation}
 \fr{\partial V}{\partial\phi_i}\left(\phi_1=\langle\phi_1\rangle,\;
\phi_2=\langle\phi_2\rangle\right)=0\;\mbox{ with }
\langle\phi_1\rangle =\fr{1}{\sqrt{2}}\left(\begin{array}{c} 0\\
v_1\end{array}\right), \; \langle\phi_2\rangle
=\fr{1}{\sqrt{2}}\left(\begin{array}{c}0
\\[5mm] v_2 e^{i\xi}\end{array}\right). \label{veveq}
\end{equation}
The commonly used parameterization of v.e.v.'s is
\begin{equation}
v_1=v\cos\beta\,,\;\; v_2=v\sin\beta\,,\quad
 \beta\in (0,\,\fr{\pi}{2})\,.
\label{vevdef}
\end{equation}
 \end{subequations}
The SM constraint $v=\left(G_F\sqrt{2}\right)^{-1/2}=246\mbox{
GeV}$ limits  the parameters of potential.

$\blacksquare$ Let us express coefficients $m_{ij}^2$ in terms of
$\lambda_i$ and v.e.v.'s, i.e. find them as solutions of
eq-s~(\ref{veveq}). It is useful to make it in two steps. We start
with the case $m_{12}=0$ (with exact $Z_2$ symmetry), which gives
$\hat{m}_{11}^2$ and $\hat{m}_{22}^2$ and equation
$Im\left(\lambda_5e^{2i\xi}\right) =0$  for the phase $\xi$. Next,
we find $m_{ij}^2-\hat{m}_{ij}^2$. This procedure contains
ambiguity, parameterized by single additional parameter denoted as
$\mu$:
 \begin{subequations}
 \begin{equation}\begin{array}{c} {\cal
M}(\phi)=-\fr{1}{2}[\hat{m}^2_{11}(\phi_1^\dagger\phi_1)+
\hat{m}^2_{22}(\phi_2^\dagger\phi_2)] +\fr{\mu^2}{2v^2}
\left(v_2e^{-i\xi}\phi_1^\dagger-v_1\phi_2^\dagger\right)
\left(v_2e^{i\xi}\phi_1-v_1\phi_2\right)\\[4mm] +
iv_1v_2Im\left(\lambda_5e^{2i\xi}\right)[e^{-{i\xi}}\phi_1^\dagger\phi_2
-h.c.]\,;\\[4mm]
 \hat{m}^2_{11}=\lambda_1v_1^2+\lambda_{345}v_2^2
\,,\;\; \hat{m}_{22}^2=\lambda_2v_2^2+\lambda_{345}v_1^2\,,\;\;
 (\lambda_{345}= \lambda_3+\lambda_4+
Re\left(\lambda_5e^{2i\xi}\right))\,,\\[4mm] \Rightarrow
m_{11}^2=\hat{m}_{11}^2+\fr{\mu^2v_2^2}{v^2},\;
 m_{22}^2=\hat{m}_{22}^2+\fr{\mu^2v_1^2}{v^2},\;
 m_{12}^2=\left[-\fr{\mu^2}{v^2}+i
 Im(\lambda_5e^{2i\xi})\right]v_1v_2e^{-i\xi}.
 \end{array}\label{solution}
 \end{equation}

These relations  present the explicit form of eq-s (\ref{veveq})
for obtaining two v.e.v.'s $v_1$, $v_2$ (or $v$ and $\tan\beta$)
and their relative phase $\xi$ via $m_{ij}$, $\lambda_i$ and
$\mu$. The latter equation of (\ref{solution}) gives $\xi$, it can
be also written as
 \begin{equation}
Im(m_{12}^2e^{i\xi})=Im (\lambda_5e^{2i\xi})v_1v_2\,.
  \label{constr}\end{equation}
\end{subequations}

$\blacksquare$ The potential (\ref{baspot}) is invariant under the
rotations
 \begin{subequations}
 \label{phinv}
 \begin{equation}
\phi_i\to e^{i\rho_i}\phi_i\;\; (i=1,2),\;\;\lambda_5\to
\lambda_5e^{2i(\rho_2-\rho_1)}\,,\;\; m_{12}^2\to
m_{12}^2e^{i(\rho_2-\rho_1)}\label{phitrans}
 \end{equation}
with $\lambda_{1-4}\,,m_{11}\,,m_{22}$ being constant. At this
transformation the phase conventions (\ref{vev}) for v.e.v.'s are
shifted as\  $\xi\to\xi+\rho_1- \rho_2$ and the quantities
$\lambda_5e^{2i\xi}$, $m_{12}^2e^{i\xi}$, $\lambda_{1-4}$ are
invariant.

The Yukawa term is invariant under this transformation if in
addition to (\ref{phitrans}) fermion fields transform as
\begin{equation}
q_{iL}\to q_{iL}e^{i\theta_i}\,,\;\; U_{iR}\to
U_{iR}e^{i(\theta_i-\rho_{iu})}\,,\;\; D_{iR}\to
D_{iR}e^{i(\theta_i-\rho_{id})}\; \mbox{ with }iu,\,id =1\mbox{ or
}2\label{fermtrans} \end{equation}
 \end{subequations}
and off-diagonal in family index Yukawa couplings  transform like
eq.~(\ref{phitrans}).

Therefore, there is a family of potentials which give the same
physics but differ in values of some parameters -- {\bf phase
family}. The choice of definite term within the family is similar
to gauge fixing for gauge fields. We denote this choice as {\em
phase gauge}. In particular, one can consider the {\em vacuum CP
symmetric phase gauge}, in which there is no "spontaneous
violation of CP symmetry of vacuum", with $\xi=0$. In other words,
the invariance mentioned above allows to eliminate phase $\xi$
from equations without changing the physically explicit CP
violation (given by the mixing of scalar and pseudoscalar
components of $\phi_i$ in the observable Higgs fields). To obtain
this representation, we start with an arbitrary form of potential,
and
 \begin{subequations}\label{xi=0}
 \begin{equation}\begin{array}{l}
 \bullet\mbox{ determine } v_i \mbox{ and } \xi \mbox{ via
equations (\ref{solution}) ,}\\[2mm]
 \bullet\mbox{ change } m_{12}^2\to
m^2_{12,0}= m_{12}^2e^{i\xi}\,,\;\;
\lambda_5\to\lambda_{5,0}=\lambda_5e^{2i\xi} \mbox{ in potential }.
\end{array}\label{xi0}
 \end{equation}
In accordance with eq.~(\ref{constr}), new values $m_{12}^2$ and
$\lambda_5$ will be constrained by
 \begin{equation}
Im(m_{12,0}^2)= Im(\lambda_{5,0})v_1v_2\quad\mbox{ (at } \xi=0)\,.
\label{constraint}
 \end{equation}
(In particular, for the $Z_2$ symmetric case we have $m_{12}=0$
and the quantity $\lambda_{5,0}$ is real.)

We use below this phase gauge, in which the general Higgs
potential (without $V^4_m$ term) adopts the form
 \begin{equation}\begin{array}{c}
V=\fr{\lambda_1}{2}(\phi_1^\dagger\phi_1-\fr{v_1^2}{2})^2
+\fr{\lambda_2}{2}(\phi_2^\dagger\phi_2-\fr{v_2^2}{2})^2+
\lambda_3(\phi_1^\dagger\phi_1-\fr{v_1^2}{2})
(\phi_2^\dagger\phi_2-\fr{v_2^2}{2})\\[4mm]
+\lambda_4(\phi_1^\dagger\phi_2) (\phi_2^\dagger\phi_1)
 +\left[\fr{\lambda_{5,0}}{2}(\phi_1^\dagger\phi_2)^2
+h.c.\right]
-\fr{[\lambda_4+Re(\lambda_{5,0})]
[(\phi_1^\dagger\phi_1)v_2^2+(\phi_2^\dagger\phi_2)v_1^2]}{2}\\[4mm]
+\fr{\mu^2}{2v^2}(v_2 \phi^\dagger_1 - v_1\phi^\dagger_2) (v_2
\phi_1 - v_1\phi_2)- iIm(\lambda_{5,0}) v_1v_2
\left(\phi^\dagger_1 \phi_2-h.c.\right) .
\end{array}\label{bas21}
 \end{equation}
\end{subequations}

$\blacksquare$ The parameters of Higgs potential are limited by
two types of conditions. First, the potential must be positive at
large quasi--classical values of fields $|\phi_i|$ to have  stable
vacuum ({\em positivity constraints}). Constraints of the other
type are related to the limits of the tree approximation
applicability. It is correct when the radiative corrections (RC)
to observable quantities are small. In this respect the {\em
perturbativity (or unitarity) limitations} are considered. These
constraints limit only our analysis. Therefore, it can happen that
the RC for all observable quantities are small while RC to the
parameters of heavy (non observed to the moment) Higgs bosons can
be large.

\subsection{ Physical sector. CP violation.}

The standard decomposition of fields $\phi_i$ in terms of physical
fields is
\begin{equation}
\phi_1=\left(\begin{array}{c} \varphi_1^+ \\[2mm]
\fr{1}{\sqrt{2}}(v_1+\eta_1+i\chi_1)
\end{array}\right), \qquad
\phi_2=\left(\begin{array}{c} \varphi_2^+ \\[2mm]
\fr{1}{\sqrt{2}}(v_2e^{i\xi}+\eta_2+i\chi_2)
\end{array}\right).\label{videf}
\end{equation}

The combinations $G^0= \chi_1\cos\beta + \chi_2\sin\beta$ and
$G^\pm=\phi_1^\pm\cos\beta +\phi_2^\pm\sin\beta$ describe
Goldstone bosons.  In the CP conserving case (at
$Im(m_{12,0}^2)=0$) the physical Higgs sector of 2HDM contains two
charged Higgs bosons $H^\pm$, one CP-odd (pseudoscalar) $A$ and
two CP-even scalars $h$ and $H$ with $M_H>M_h$
 \begin{equation}\begin{array}{c}
A=- \chi_1\,\sin\beta+\chi_2\,\cos\beta\,,\quad
H^\pm=-\varphi_1^\pm\,\sin\beta+\varphi_2^\pm\,\cos\beta\,,\\[4mm]
H=\eta_1\,\cos\alpha +\eta_2\,\sin\alpha\,,\quad
h=-\eta_1\,\sin\alpha+\eta_2\,\cos\alpha\,; \quad
\alpha\in(-\pi/2,\,\pi/2)\,.\end{array} \label{physfields}
 \end{equation}

In the general \noCP case we have the same charged Higgs bosons
with mass $M_{H^\pm}^2=\mu^2-(\lambda_4+Re(\lambda_{5,0}))v^2/2$
and three neutral scalar states $h_1$, $h_2$, $h_3$ having no
definite CP parity (with convention $M_{h_3},\, M_{h_2}\ge
M_{h_1}$). These $M_{h_i}^2$ are obtained from $\eta_1$, $\eta_2$
and $A$ by diagonalization of the mass matrix $M$
 \begin{equation}\begin{array}{c} M=\left(\begin{array}{ccc}
 M_{11}\;\;&M_{12}
& \delta\sin\beta\\[4mm] M_{12}& M_{22} & \delta\cos\beta\\[4mm]
\delta\sin\beta\;\;& \delta\cos\beta
 \;\;&
 \mu^2-Re(\lambda_{5,0})v^2\end{array}\right)\\[4mm]
 \\
 M_{11}=\mu^2\sin^2\beta+\lambda_1v^2\cos^2\beta,\;\;
M_{22}= \mu^2\cos^2\beta+\lambda_2v^2\sin^2\beta,\\[4mm]
 M_{12}=(\lambda_{345}v^2 -\mu^2)\fr{sin2\beta}{2},\;\;
 \lambda_{345}= \lambda_3+\lambda_4+ Re\lambda_{5,0}\,;\quad
\delta= \fr{Im(\lambda_{5,0})v^2}{2}\,.
 \end{array}\label{MM}\end{equation}

These states are determined with the aid of unitary transition
matrix $R$, written through Euler angles $\alpha_i$
($c_i=\cos\alpha_i, \;\;s_i=\sin\alpha_i$):
 \begin{equation}\begin{array}{c}
\left(\begin{array}{c}h_1\\h_2\\h_3\end{array}\right)
=R\left(\begin{array}{c}\eta_1\\ \eta_2\\A
\end{array}\right), \;\; R= \left(\begin{array}{ccc} -s_1& c_1c_2
&-c_1s_2 \\ -c_1s_3 &\;s_2c_3-s_1c_2s_3\;&s_1s_2s_3+c_2c_3\\
c_1c_3&s_1c_2c_3+s_2s_3&c_2s_3-s_1s_2c_3
\end{array}\right).\\[8mm]
\left(\mbox{At }\;\alpha_2,\; \alpha_3 \to 0\Rightarrow
\alpha_1\to \alpha,\;\;h_1\to h, \;\;h_2\to H,\;\; h_3\to
A\right)\,.
\end{array}\label{mixing}
 \end{equation}

$\bullet$ {\em In the CP conserving case} the masses of neutral
Higgs particles and mixing angle $\alpha$ are obtained by
diagonalization of mass matrix $M$ (\ref{MM}) with $\delta=0$:
 \begin{subequations}\label{masses}
\begin{equation}\begin{array}{c}\begin{array}{c}
M_A^2=\mu^2-Re(\lambda_{5,0}) v^2, \quad
M_{H^\pm}^2=\mu^2-\fr{1}{2}(Re(\lambda_{5,0})+\lambda_4)v^2,\\
[4mm]
 M_{h,H}^2=\fr{1}{2}\left(M_{11} +M_{22}\mp R\right),\quad
 R=\sqrt{(M_{11}-M_{22})^2+4M_{12}^2},\\[4mm]
 M_H^2+M_h^2=\mu^2+\fr{1}{2}[(\lambda_1+\lambda_2)+
 (\lambda_1-\lambda_2)\cos2\beta]v^2\,.
\end{array}\end{array}\label{difmas}
 \end{equation}
These equations can be accompanied by useful relations:
\begin{equation}
 \sin2\alpha=\fr{2M_{12}}{R},\quad
 \cos2\alpha=\fr{M_{11}-M_{22}}{R}\,,\;\;
M_H^2-M_h^2=
\fr{\sin2\beta}{\sin2\alpha}\left(-\mu^2+\lambda_{345}v^2\right)\,.
\label{massadd} \end{equation}

$\bullet$ {\em In the case of weak ($\phi_1,\,\phi_2$) mixing},
i.e. small value of $m_{12}$ and respectively $\delta$, one can
obtain the neutral Higgs boson masses in the form of corrections
$\sim\delta^2$ to the expressions (\ref{difmas}),
 \begin{equation}\begin{array}{c}
M_1^2=M_h^2-\fr{\delta^2}{2(M_A^2-M_h^2)}
 (1+t)\,,\quad
 M_2^2=M_H^2-\fr{\delta^2}{2(M_A^2-M_H^2)}(1-t)
 \,,\\[3mm]
M_3^2=M_A^2+M_H^2+M_h^2-M_1^2-M_2^2\,,\quad  t=
\fr{(M_{11}-M_{22})\cos2\beta- M_{12}\sin2\beta}{M_H^2-M_h^2}\,.
\end{array}\label{masspert}
 \end{equation}
Similarly, one can easily obtain {\em small} mixing angles
$\alpha_2$ and $\alpha_3$ just as  $\alpha_1-\alpha$ in this case.
 \end{subequations}

$\blacksquare$ {\bf Some couplings}. The couplings of Higgs
particles to vector bosons are written via elements of transition
matrix $R$ (\ref{mixing}):
 \begin{equation}
 \chi_V^{h_i}=s_\beta R_{i2}+c_\beta R_{i1}\,.\label{chiV}
\end{equation}

As well as other trilinear couplings, the coupling of charged
Higgs boson to the neutral Higgs boson is written via $\lambda_i$
and $v_i$, without $\mu$. Hereafter it is useful to present this
coupling in terms of observable masses and $\mu$, in units of the
coupling of the Higgs particle $\phi=h$ (or $H$) to an arbitrary
scalar particle with mass equal to $M_{H^{\pm}}$ added to the SM.
For the CP conserving case that is
\begin{equation}
\chi_{H^\pm}^\phi \equiv-\fr{vg_{\phi H^+H^-}}{2M_{H^\pm}^2}
=\left(1-\fr{M_\phi^2}{2M_{H^\pm}^2}\right)\sin(\beta-\alpha)
+\fr{(M_\phi^2-\mu^2)\cos(\beta+\alpha)}
{M_{H^\pm}^2\sin2\beta}\,. \label{b2d2}
\end{equation}

\subsection{ Different scenarios in 2HDM, natural range of
parameters.} \label{difscen}

The widely discussed scenario for Higgs sector is that there
exists one light Higgs boson $h_1$ which is near discovery reach
of present accelerators, and other possible Higgs bosons are very
heavy. In accordance with eqs.~(\ref{masses}), it can takes place
if

$\bf 1.$ {\bf\boldmath $|m_{12}|$ is small ($\mu\lesssim v$)},
i.e. $Z_2$ symmetry is precise or weakly broken. Some Higgs bosons
can be heavy due to large values of couplings $\lambda_i$, their
values are limited from above by perturbativity limitation for
$\lambda_i$ at the level $ 4\pi v\approx  3$ TeV. In particular,
if additionally CP is conserved and masses of $H$, $A$, $H^\pm$
are about 800 GeV, for the parameters of perturbation theory we
have $|\lambda_5|/(4\pi)^2$, $\lambda_4|/(4\pi)^2$,
$|\lambda_{345}|/(4\pi)^2 \approx 0.07$. These values lie within
the perturbativity domain.

$\bf 2.$ {\bf\boldmath $\mu$ is large ($ \mu^2\gg \lambda_i
v^2\,,M_h^2$)}, i.e. $Z_2$ symmetry is strongly broken. In this
case high values of masses are given by high value of parameter
$\mu$ even at relatively small Higgs boson self--couplings
$\lambda_i$. Besides, in this case due to eqs.~(\ref{masses}),
\begin{Itemize}
\item Heavy Higgs bosons $H$, $A$ (or $h_2$, $h_3$) and $H^\pm$
are almost degenerate in their masses.

\item Comparing the last equations (\ref{difmas}) and (\ref{massadd}),
in the CP conserving case we have $\sin2\alpha\approx
-\sin2\beta$. Since $\beta\in(0,\pi/2)$ and
$\alpha\in(-\pi/2,\pi/2)$, it gives two cases:
\begin{Itemize}
\item[A.] $2\alpha\approx 2\beta-\pi\Rightarrow \sin(\beta-\alpha)
\approx 1$. It makes coupling constants of the lightest Higgs $h$
to the gauge bosons and quarks (in Model II) be close to their SM
values. This very case is treated usually as {\bf decoupling
limit} \cite{Haber}.
\item[B.] $\alpha\approx -\beta$. This solution gives couplings of
Higgs boson to gauge fields and quarks which are generally far
from their SM values.
\end{Itemize}
 \end{Itemize}

$\bullet$ The eq.~(\ref{MM}) shows explicitly that the \noCP can
take place only if $Im(m_{12,0}^2)\neq 0$, i.e. in the case when
$\phi_1$ and $\phi_2$ are mixed. The observed weak \noCP means
relatively small value of $Im(m_{12}^2)$ in the specific "vacuum
CP conserving phase gauge (\ref{phinv})". This condition is {\em
natural}, i.e. it can be formulated independently on phase gauge
if $|m_{12}|$ is also small, and it looks as {\em unnatural phase
gauge dependent condition} if $|m_{12}|$ is not small  (for the
decoupling limit).  The similar discussion of FCNC is given below.
\vspace{2mm}

\noindent
 \begin{fmpage}{0.99\textwidth}
Therefore, {\bf\boldmath a \underline{weak} \noCP and FCNC effects
{\em naturally} correspond to the case $|m_{12}|\ll v$, i.e. the
case of {\underline{weak}} $Z_2$ symmetry breaking (first
opportunity above), and they are {\em unnatural}\  for the the
decoupling limit.}
 \end{fmpage}
\vspace{3mm}

The discussed opportunities are summarized in the Table
\ref{tabnat}.

\begin{table}[hbt]
\begin{tabular}{|c|c|c|c|c|}
\hline &\multicolumn{2}{|c|}{basic relations in}& {small \noCP}
&specific\\
 & ${\cal L}$ (\ref{baspot})&${\cal L}$ (\ref{bas21})&and FCNC& features\\\hline
 weakly&$|m_{12}^2|\lesssim v^2$&$\mu^2\lesssim
v^2$,& natural&$M_{2,3},\,M_{H^\pm}<3$ TeV,\\
 broken && $Im\left(m_{12,0}^2\right)\lesssim v^2$&&
$\;\;\chi_{H^\pm}\sim 1$\\
 $Z_2$ symmetry&&&&\\ \hline
 decoupling&$Re\left(m_{12}^2e^{i\xi}\right)\gg v^2$,
&$\mu^2\gg v^2$,&unnatural &$M_{2,3}\approx M_{H^\pm}$,
\\
 limit&$Im\left(m_{12}^2e^{i\xi}\right)$&
 $Im\left(m_{12,0}^2\right)$&&$\;\;\chi_{H^\pm}\ll 1$\\
\hline
 \end{tabular}
\caption{\it Scenarios of 2HDM considered as natural in this paper
and in some other papers. } \label{tabnat}
\end{table}

\subsection{ Coupling to fermions (Yukawa interaction)}

To avoid ($\phi_1,\,\phi_2$) mixing at the one loop level, each
type of quarks or leptons should acquire mass via only one Higgs
field
 \begin{equation}\begin{array}{c} { \cal L}_Y^0=\sum \limits _i
g_i^d\bar{q}_{iL}\phi_{id} D_{iR}+\sum \limits_i g_i^u
\bar{q}_{iL} \phi_{iu} U_{iR}+h.c.+ \mbox{lepton items} \\[2mm]
\mbox{with some of } iu, id,i \ell =1\;\mbox{ and other }=2 \, .
\end{array}\label{Yuk0}
 \end{equation}
Two variants of this type are widely discussed in the literature
(cf. \cite{Hunter}), these are {\it Model I } with $id=iu=i\ell
=1$ and {\it Model II } considered below.

If Higgs potential contains ($\phi_1,\,\phi_2$) mixed terms, there
are no grounds to have Yukawa interaction in the form
(\ref{Yuk0}). Each fermion can be coupled to both Higgs bosons,
and Yukawa interaction  can be even off-diagonal in family index,
giving FCNC. Assuming violation of $Z_2$ symmetry to be weak, we
can consider these off--diagonal terms  to be small. For Model I
similar approach was developed in \cite{wu}. We neglect these
small $Z_2$ symmetry violating Yukawa interactions in our
discussion of Higgs sector itself.

$\blacksquare$ {\bf Model II (2HDM~(II)).}

We consider in detail the  Model II in which the couplings to
fermions are similar to those in MSSM. In this model, the
fundamental scalar field $\phi_1$ couples to $u$-type quarks,
while $\phi_2$ -- to $d$-type quarks and charged leptons (we
assume neutrinos to be massless),
 \begin{equation}
{ \cal L}_Y^0=\sum \limits_i g_i^d \bar{q}_{iL} \phi_2 D_{iR}+\sum
\limits _i g_i^u \bar{q}_{iL} \phi_1 U_{iR}+\sum \limits _i g_i^{
\ell} \bar { \ell}_{iL} \phi_2 E_{iR}+h.c. \,  \label{YukII}
 \end{equation}
Here the ratios, relative to the SM values, of the direct coupling
constants of the Higgs boson $h_i$ to the gauge bosons $V=W$ or
$Z$, to up and down quarks (\ref{Eq:chi-def}) ({\em basic
couplings}) are given by elements of transition matrix $R$
(\ref{mixing}) as (\ref{chiV})
 \begin{equation}
 \chi^i_u=\fr{R_{i2}- i\gamma^5 \cos\beta R_{i3}}{\sin\beta}\,,\quad
 \chi^i_d=\fr{R_{i1}- i\gamma^5 \sin\beta R_{i3}}{\cos\beta}\,.
 \label{mixcoupl1}
 \end{equation}

$\Box$ It is also useful to write coupling constants of neutral
scalars $\phi=h$ or $H$ to charged Higgs boson (\ref{b2d2}) via
couplings of these scalars to fermions. In the CP conserving case
\begin{equation}
\chi_{H^\pm}^\phi \equiv-\fr{vg_{\phi H^+H^-}}{2M_{H^\pm}^2}
=\left(1-\fr{M_\phi^2}{2M_{H^\pm}^2}\right)\chi_V^{\phi}
+\fr{M_\phi^2-\mu^2}{2M_{H^\pm}^2}
(\chi_u^{\phi}+\chi_d^{\phi})\,,\quad (\phi=h\mbox{ or }H)\,.
\label{b2d3}
\end{equation}

The difference between  two scenarios for mass generation of
sect.~\ref{difscen}\ \  influences strongly this coupling. It is
clearly seen that for the lightest Higgs boson in the first
scenario (with natural \noCP and FCNC) $\chi_{H^\pm}\approx
\chi_V\sim 1$ while for the decoupling limit $\chi_{H^\pm}\ll 1$.
The measurement of the Higgs boson production in \ggam\ collisions
can distinguish these two mechanisms in the experiments at Photon
Colliders \cite{GKO1}. (Note that direct measurements of separate
Higgs self--couplings may be difficult).

$\blacksquare$ {\bf Pattern relation and sum rules}.

$\bullet$ The unitarity of matrix $R$ allows to obtain a simple
relation on couplings of one Higgs particle to gauge bosons and
quarks, which is very useful at the phenomenological analysis, --
{\em pattern relation}\fn{In the CP conserving case for $\phi=h$,
$H$ and $A$ three observable $\chi_i$ can be expressed via {\it
two} angles, $\beta$ and $\alpha$. Therefore $\chi_V$, $\chi_u$
and $\chi_d$ cannot be independent, that gives (\ref{2hdmrel}).}
\cite{GKO1,GKO3}, having the same form for each Higgs boson $h_i$:
\begin{equation}
(\chi_u +\chi_d)\chi_V=1+\chi_u \chi_d. \label{2hdmrel}
 \end{equation}
It is also useful to express $\tan\beta$ via these  couplings:
\begin{equation}
\tan^2\beta={\fr{(\chi_V-\chi_d)^\dagger}{\chi_u-\chi_V}}\,.
\label{Eq:tan-beta}
\end{equation}

$\bullet$  It can be obtained that unitarity of the matrix $R$
brings also the {\em sum rules} for each neutral Higgs boson
\cite{GUN}:
 \begin{equation}
|\chi_u|^2\sin^2\beta+|\chi_d|^2\cos^2\beta=1\,.\label{srules}
 \end{equation}
These sum rules guarantee that the production cross sections for
each neutral Higgs boson of 2HDM cannot be lower than that for the
SM Higgs boson with the same mass at least in one commonly
discussed process \cite{GUN}.

\section{Standard Model-like scenario}\label{secSMlike}

We now consider the following scenario, referred below to as the
{\it SM-like scenario}. It is defined by the following
criteria\fn{\it\normalsize This part is based on the papers
\cite{GKO3}.}:

{\bf 1. One Higgs boson will be discovered} (with mass above
today's limit for the SM Higgs boson, 115~GeV). {\em It can either
be the Higgs boson of the SM or one of several neutral scalars of
another model, such as 2HDM.}

{\bf 2.  No other Higgs boson will be discovered.} That means, that
other possible scalars are either\\
 $\Box$  weakly coupled to the $Z$ and $W$ bosons, gluons and
quarks, or\\
 $\Box$  heavy enough to escape (direct or indirect)
observation, e.g. $ M_{H^{\pm}}> {\cal O}(800\mbox{ GeV})$.

{\bf\boldmath 3. Any other new particles that may exist are
heavier than the discovery limits of LHC and the $e^+e^-$ Linear
Collider.}

{\bf\boldmath 4. The measured decay widths of the observed Higgs
boson (or the squared coupling constants) to other particles,
$\Gamma_i^{\rm exp}$, will be in agreement with their SM values
$\Gamma_i^{\rm SM}$ within the to-date precision $\delta_i$}, i.e.
 \begin{subequations}
\label{widthest}
 \begin{equation}
\left|\frac{\Gamma_i^{\rm exp}}{\Gamma_i^{\rm SM}}-1\right|
\lesssim\delta_i\ll 1.\label{width1}
 \end{equation}
For the coupling constants themselves in terms of (\ref{chidef}),
the eq.~(\ref{width1}) means that
 \begin{equation}
\chi_i=\pm (1-\epsilon_i)\;\mbox{ with }
|\epsilon_i|\le\delta_i\,.\label{width2}
 \end{equation}
 \end{subequations}

\subsection{SM-like realizations in the 2HDM~(II)}

Even in the simplest extension of SM, the 2HDM, the SM-like
scenario can be realized in many regions in the parameter space.
We consider in detail the CP conserving case. There are two
classes of solutions denoted $A_{\phi\pm}$ and $B_{\phi\pm q}$.
Here the first subscript labels the observed Higgs boson and
second subscript labels the sign of $\chi_V^\phi$. For the
solutions $A_{H\pm}$, $B_{H\pm q}$ the analysis includes also the
veto for the discovery of the lightest Higgs boson in the
associated production with $t$ or $b$ quarks. More detailed
analysis is available, for example, in \cite{GKO3}.

{\em For solutions $A_{\phi\pm}$} basic couplings of observed
Higgs boson are approximately identical, $\chi_V^\phi \approx
\chi_u^\phi\approx \chi_d^\phi\approx \pm 1$. For example, the
solution $A_{H-}$ is that with the observed Higgs boson being the
heaviest one, $H$, and with $\chi_V^H\approx -1$. (The decoupling
limit can be realized for solution $A_{h+}$.) The exact solutions
of these types $|\chi_i|=1$ can also be realized.

{\em For solutions $B_{\phi \pm q}$} some of basic $\chi_i\approx
1$ but other $\chi_j\approx -1$. The third subscript $q=d,u$
denotes the type of quark whose coupling with the observed Higgs
boson is of opposite sign as compared to the gauge boson coupling,
$\chi_V$. (The solutions with $-\chi_V \approx \chi_d\approx
\chi_u\approx \pm 1$ cannot be realized.) The exact solutions of
these types $|\chi_i|=1$ cannot be realized, the conditions
$\epsilon_V\neq 0$ and $\tan\beta\gg 1$ or $\ll 1$ are necessary
in these cases. Note that these solutions cannot be realized at
the decoupling limit, for all of them there should be $\mu\lesssim
v$.

The allowed realizations of the SM-like scenario in the CP
conserving case in 2HDM (II) are listed in the Table~\ref{Tab2}.
The numbers in the table correspond to (1) the observed Higgs
boson having the mass 115-180 GeV, considering higher masses of
this boson for completeness; (2) usage of the anticipated
inaccuracies for the Higgs boson couplings to quarks and gauge
bosons at the $e^+e^-$ Linear Collider \cite{TESLATDR}.

\begin{table}[bht]
\begin{center}
\begin{tabular}{||c|c|c|c|c|c|c||}
\noalign{\vspace{-8.5pt}} \hline\hline
&&observed&&\multicolumn{2}{|c|}{}&\\
 type &notation &Higgs
&$\chi_V$&\multicolumn{2}{|c|}{$\tan\beta$}&constraint\\
&&boson&&\multicolumn{2}{|c|}{}&\\ \hline
  &$A_{h+}$&h
&$\approx+1$&&$\lessgtr 1$&\\ \cline{2-4}\cline{6-6}
 $A_{\phi\pm}$ &$A_{H+}$&H
&$\approx+1$ &&$\lessgtr 1$&\\ \cline{2-4}\cline{6-6}
$\chi_V\approx\chi_u\approx\chi_d$&$A_{h-}$& h&$\approx
-1$&$\sqrt{\left|\fr{\epsilon_d}{\epsilon_u}\right|}$ & $\ll
1$&$\epsilon_V=-\fr{\epsilon_u\epsilon_d}{2}$\\\cline{2-4}\cline{6-6}
 &$A_{H-}$&H&$\approx -
1$&&$\gg 1$&\\ \hline \vspace{-4mm}&&&&\multicolumn{2}{|c|}{} &\\
 $B_{\phi\pm d}:$&$B_{h+d}$&h&$\approx +1$&
\multicolumn{2}{|c|}{$\sqrt{\fr{2}{\epsilon_V}}\gtrsim 10$} &
$\epsilon_u=-\fr{\epsilon_V\epsilon_d}{2}$\\\cline{2-4}
 $\chi_V\approx\chi_u\approx-\chi_d$ &$B_{H\pm d}$&H&$\approx \pm 1$&
\multicolumn{2}{|c|}{} &
\\ \hline
\vspace{-4mm}  & &&&\multicolumn{2}{|c|}{} &\\
 $B_{\phi\pm u}:$
&$B_{h\pm u}$&h&$\approx \pm 1$&
\multicolumn{2}{|c|}{$\sqrt{\fr{\epsilon_V}{2}}\lesssim 0.1$}&
$\epsilon_d=-\fr{\epsilon_V\epsilon_u}{2}$\\\cline{2-4}
 $\chi_V\approx\chi_d\approx-\chi_u$  &$B_{H+u}$&H&$\approx
+1$& \multicolumn{2}{|c|}{}&
\\\hline
 \multicolumn{7}{||c||}{\vspace{-4mm}}\\ \multicolumn{7}{||c||}{
$\chi_i=\fr{g_i}{g_i^{SM}}=\pm(1-\epsilon_i)$ with}\\
\multicolumn{7}{||c||}{ $i=V(\equiv Z,\,W)$ or $i=u(\equiv t,\,
c)$ or $i=d,\ell(\equiv b,\, \tau);\quad\epsilon_V>0$,
$\epsilon_u\epsilon_d<0$ .}\\\hline\hline
\end{tabular}
\vspace{3mm} \caption{Allowed realizations of SM-like scenario in
the 2HDM~(II)}
\end{center}
\label{Tab2}
\end{table}

\section{Resolving SM-like scenarios via Higgs boson production at
a Photon Collider}

Study of Higgs-boson couplings with photons ($h\gamma\gamma$
and $hZ\gamma$) is a very promising tool for resolving the
models of New Physics by the following reasons.

{\bf$\bullet$ These couplings are absent in the SM at tree level,
appearing only at the loop level.} Therefore, the background for
signals of New Physics here will be relatively lower than in other
processes allowed at tree level of the SM.

{\bf$\bullet$ All fundamental charged particles contribute to
these effective couplings.} The whole structure of the theory
influences the corresponding Higgs-boson decays. Note that for the
contributions of heavy particles with masses given by the Higgs
mechanism (like in minimal SM), there is no decoupling in these
vertices.

{\boldmath\bf$\bullet$ The anticipated accuracy in the
measurements of $\Gamma(H\to\ggam)$ in the $\ggam\to h$ process at
Photon Collider is {\bf\boldmath $\sim 2$\%} with the luminosity
integral 30~fb$^{-1}$ and $M_h\le 150$ GeV} \cite{Jik},
\cite{TESLATDR}. It can be reduced to 1 \% level with the
anticipated luminosity integral about 500 fb$^{-1}$. Another
possible opportunity to study these effects is provided by the
$hZ\gamma$ interactions with the best potential for studying in
the process $e\gamma\to eh$. Certainly, possible accuracy here is
lower than in the \ggam\ channel.

\subsection{ 2HDM vs. SM}

Of course, the best place for the comparison of models is given by
the Higgs boson production in \ggam\ collisions. We calculated
these $h\ggam$ vertices in \cite{GKO1,GKO3}. Besides, we obtained
there the $hZ\gamma$ decay width deviations from SM which are
similar to those found for \ggam\ channel but lower in value. The
impression was: {\em the $hZ\gamma$ vertex is unsuitable for
resolving the models}. However, this vertex participates in
description of the process \egeh\  very far from the mass shell.
Thus, we consider here this process to check the prediction above.

The process \egeh\ is described with diagrams: ({\em i}) $\egam\to
(e\gamma^*)\gamma\otimes\gamma^*\gamma\to h$ (photon exchange);
({\em ii}) $\egam\to (eZ^*)\gamma\otimes Z^*\gamma\to h$ ($Z$
exchange); ({\em iii}) box diagrams; the latter give small
contributions. This subdivision is approximately gauge invariant.
Therefore, separate terms have physical sense \cite{BGI}.

In the total cross section of \egeh\ process the diagram with
photon exchange is dominant. At $p_\bot(e)>30$ GeV the photon and
$Z$ contributions become comparable, giving very different cross
sections for the left-hand and right-hand polarized electrons,
$\sigma_L>3\sigma_L$ \cite{BGI}. Therefore, we present only
results for $\sigma_L$ integrated over the region $p_\bot(e)>30$
GeV for $\sqrt{s_{e\gamma}}=1.5$ TeV (note that energy dependence
becomes weak at large enough energy).

We calculated the relative widths $|\chi_{\gamma\gamma}|^2$ and
the $\sigma_L(\egam\to eh)$ for all allowed realizations of
SM-like scenario in 2HDM assuming natural form of Higgs potential,
with $\mu\lesssim v$. For definiteness, we perform all
calculations for $\mu=0$, $M_{H^\pm} =800$~GeV\fn{ Since the
coupling $\chi_{H^{\pm}}$ depends linearly on $\mu^2$, $
|\chi|_{\gamma\gamma}^2
=1-R_{\gamma\gamma}\left(1-\fr{|\mu^2|}{M_{H^\pm}^2}\right)$, with
quantity $R_{\gamma\gamma}$ which is determined from
$|\chi_{\gamma\gamma}|^2$ at $\mu=0$ (and the same equation for
the ratio of $e\gamma\to eh$ cross sections). In the unnatural
case $M_{H^\pm}\approx \mu$ these measurements cannot distinguish
models, $|\chi_{\gamma\gamma}|^2=1$}. In accordance with
eq.~(\ref{b2d3}), at $M_i< 250$~GeV the contribution of the
charged Higgs boson loop varies by less than 5\% when
$M_{H^{\pm}}$ varies from 800~GeV to infinity.

In the figures with the results, ({\em{i}}) solid curves
correspond to the exact case, where all basic $|\chi_i|=1$;
({\em{ii}}) the shaded bands are derived from anticipated (in
\cite{TESLATDR}) 1~$\sigma$ bounds for the measured basic coupling
constants, $g_V$, $g_u$ and $g_d$, with additional constraints
given by the pattern relation for each solution (Table 2).

{\bf Solutions A.} A new feature of the considered widths and
cross sections in the 2HDM compared to the SM case is the
contribution from the charged Higgs boson loops. The results are
shown in Figure 1.

 \begin{figure}[htb]
\includegraphics[height=5cm,width=0.45\textwidth]{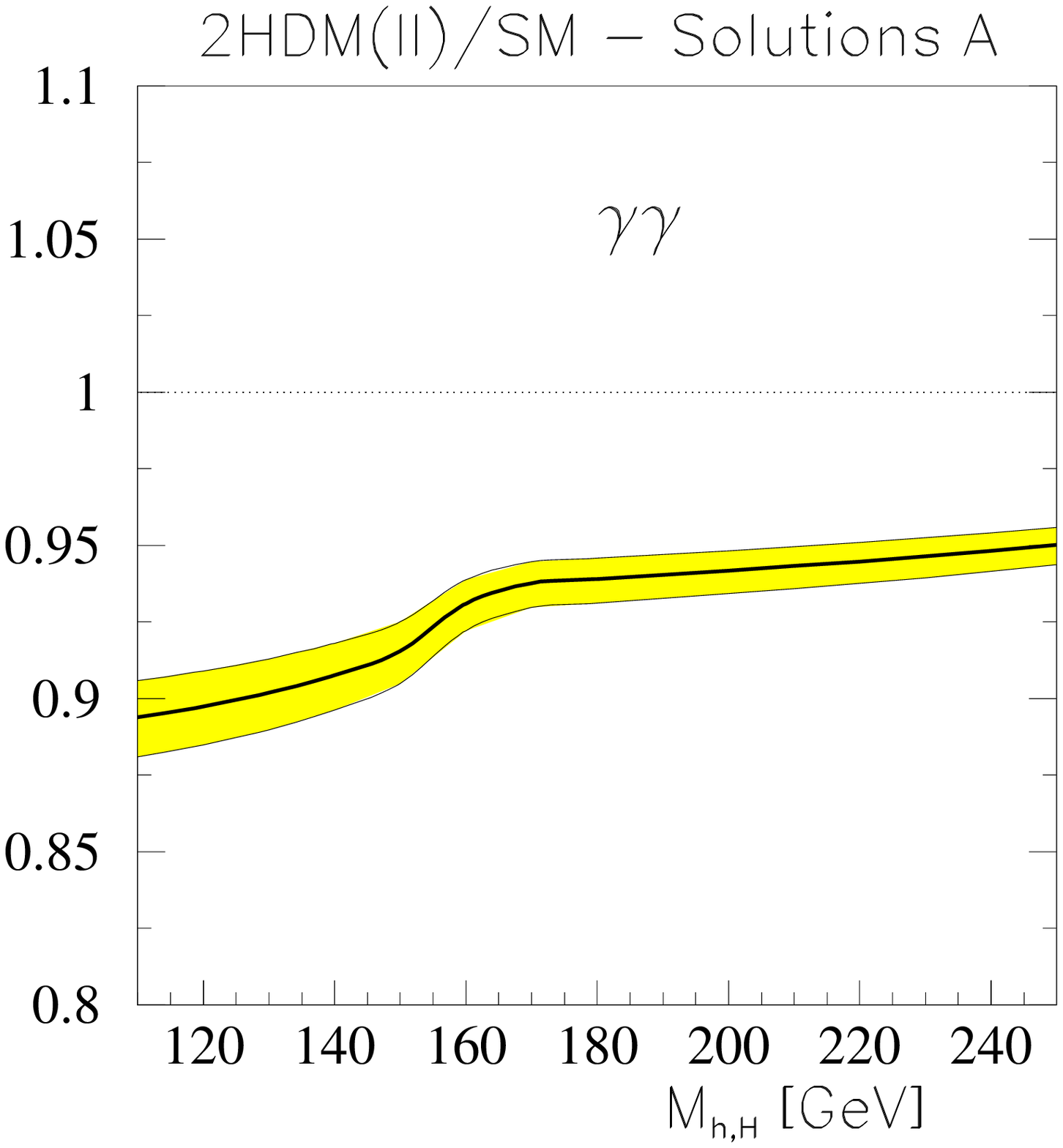}
\includegraphics[height=5cm,width=0.45\textwidth]{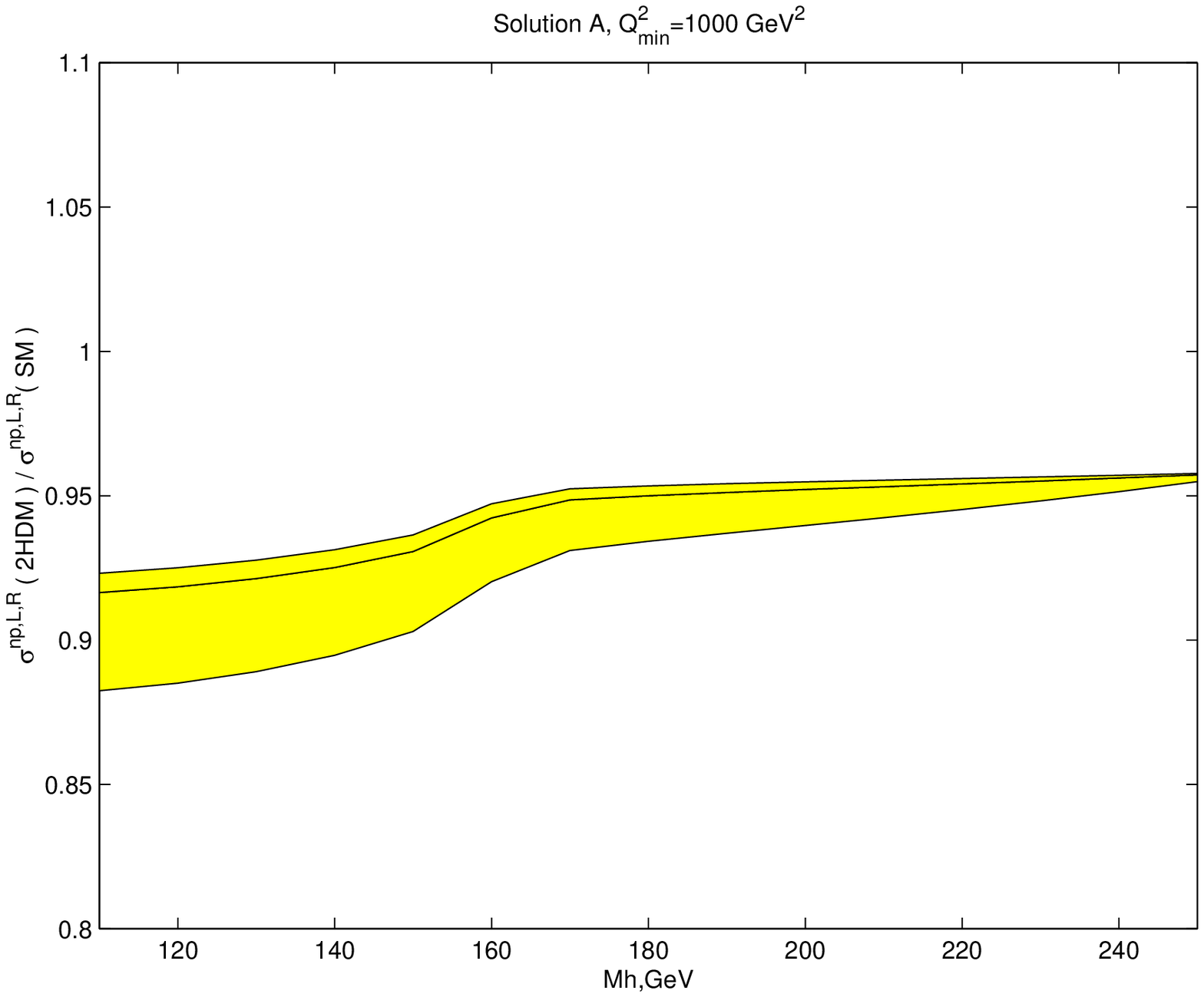}
\caption{\it Solutions A and $B_{\phi\pm d}$. The ratio of
quantities in 2HDM to their SM values. The two-photon Higgs width
-- left panel; the cross section $\sigma_L(e\gamma\to eh)$ --
right panel.}
 \end{figure}

{\bf Solutions B}. For solutions $B$ we have, by definition,
$\chi_u+\chi_d ={\cal O}(\epsilon)$. So with high accuracy
$\chi_{H^\pm} \approx \chi_V$. The results  are shown in Figure 2.
At the left panel (for \ggam) the lower curves correspond to the
solutions $B_{\phi\pm d}$ and the upper ones to the solution
$B_{h+u}$. At the right panel (for \egeh) we show the SM and
2HDM cross sections themselves.

$\bullet$ {\em For the solutions~$B_{\phi\pm d}$} the main source
of deviation from SM predictions is charged Higgs contribution.
The effect of the opposite relative sign of the $b$-quark coupling
($\chi_d\simeq-\chi_V$) as compared to that in the SM case is
negligible, since this contribution is very small itself.
Therefore, the curves for this case coincide with those for
solutions~A (Fig.~1) with only note that the exact solution cannot
be realized in this case. The result for the $\ggam\to h$
transition is also shown in the lower curve of left panel in
Fig.~2

$\bullet$ {\em For the solution $B_{h+u}$} the photon widths
increase dramatically as compared to the SM case. Here, solid
curve corresponds to the case $\chi_V=\chi_d=-\chi_u= 1$, and
$t$--quark contribution is smaller than that from $W$--boson, but
it is about 20\% from the $W$-boson one, and change of its sign
becomes essential (Fig.~2).

 \begin{figure}[htb]
\includegraphics[height=5cm,width=0.45\textwidth]{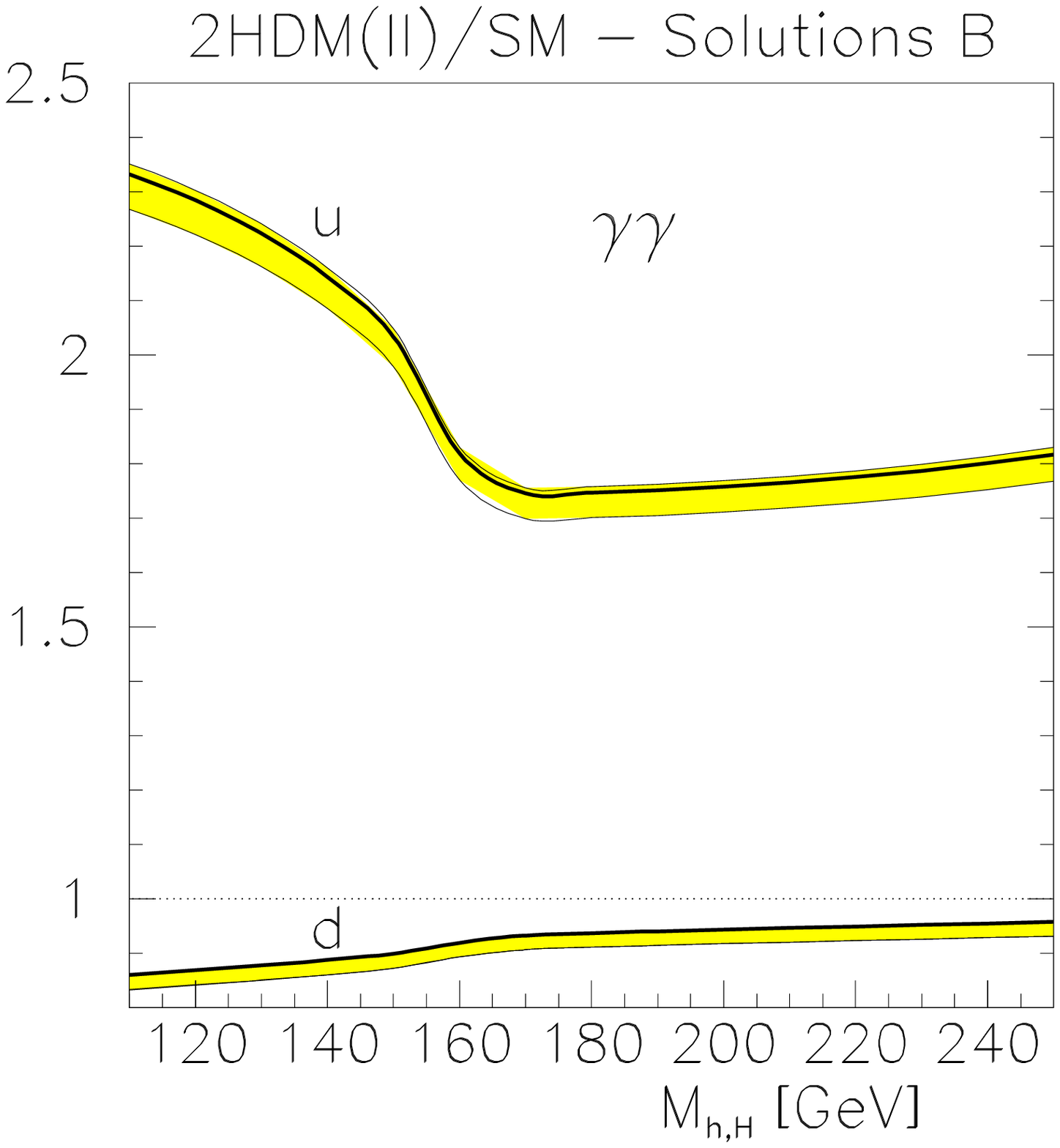}
\includegraphics[height=5cm,width=0.45\textwidth]{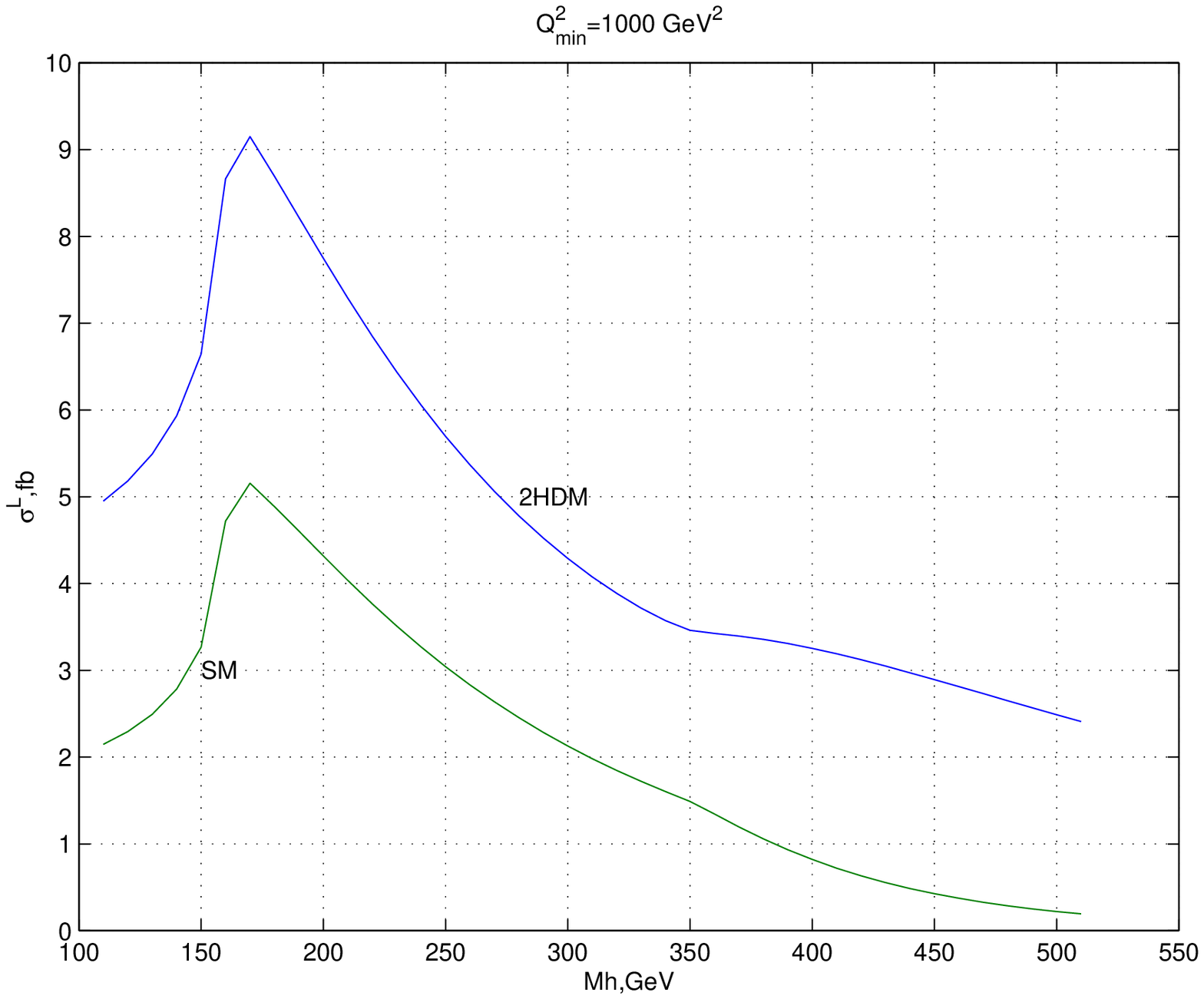}
\caption{\it Solutions B --- the ratio of two-photon widths in
2HDM to their SM values -- left panel.  The  solution $B_{h+u}$
--- the cross section $\sigma_L(e\gamma\to eh)$ in 2HDM and SM --
right panel.}
\end{figure}

\section{Conclusion and final notes }

Let us summarize main conclusions.

1. The general 2HDM, in which strong CP violation and large FCNC
effects are naturally suppressed, corresponds to small
($\phi_1,\,\phi_2$) mixing, i.e. differs substantially from the
option considered usually in context of decoupling limit.

2. Possible SM--like scenario includes the picture considered
in the description of decoupling limit and allows many other
realizations.

3. The comparison of the presented results with the anticipated
experimental uncertainty shows that the deviation of the
two-photon width from its SM value is generally large enough to
allow a reliable distinction of the natural 2HDM~(II) from the SM
at the Photon Collider. The \egeh\ process can supplement this
potential substantially, at least at $M_h<160$ GeV.

4. Solutions $B_{\phi\pm u}$ are separated well enough even for more
rough measurements and independent on possible strong CP violation
and FCNC.

5. We don't see a way in such measurements to distinguish the
cases when the observed Higgs boson is the lightest one $h$
($h_1$) or the heaviest one, $H$ ($h_2$ or $h_3$).

\vspace{8mm}

Most part of results reported here was obtained in collaboration
with M.~Krawczyk and P.~Osland. We are grateful to them for this
fruitful collaboration. We are grateful to A. Djouadi, J. Gunion,
H. Haber and M. Spira for discussions of decoupling in the 2HDM
and the MSSM, and to P.~Chankowski, W.~Hollik, I.~Ivanov, A. Pak
for valuable discussions of the parameters of the 2HDM. This
research has been supported by RFBR grants 99-02-17211 and
00-15-96691, INTAS grant 00-00679.


\begin{thebibliography}{99}

\bibitem{GKO2} I.F. Ginzburg, M. Krawczyk, P. Osland, in
preparation.

\bibitem{GKO1}
I.F. Ginzburg, M. Krawczyk, P. Osland, {\em Proc.\ 4th Int.\
Workshop on Linear Colliders, April 28-May 5, 1999; Sitges (Spain)
p. 524}, (hep-ph/9909455); I.F.~Ginzburg, Nucl. Phys. B (Proc.
Suppl.) {\bf 82}, 367 (2000); hep-ph/9907549.

\bibitem{GKO3} I.F. Ginzburg, M. Krawczyk, P. Osland,
hep-ph/0101208; \NIM {\bf A 472} (2001) 149, hep-ph/0101229;
hep-ph/0101331

\bibitem{Hunter}
J.F. Gunion, H.E. Haber, G. Kane, S. Dawson, {\em The Higgs
Hunter's Guide} (Addison-Wesley, Reading, 1990).

\bibitem{GUN} B. Grz{\c a}dkowski, J.F. Gunion, J. Kalinowski,
Phys.\ Rev.\ D {\bf 60}, 075011 (1999); Phys.\ Lett.\ {\bf B 480},
287 (2000).

\bibitem{Wolf} J. Liu, L. Wolfenstein, {\em Nucl. Phys.}\  {\bf
B289} (1987) 1.

\bibitem{Haber}
H.E. Haber,  hep-ph/9505240; hep-ph/9707213.

\bibitem{wu} Y.-L. Wu hep-ph/9404241

\bibitem{TESLATDR}  R.D. Heuer et al. TESLA Technical
Design Report,\ \  {\bf p. III} {\em DESY 2001-011, TESLA Report
2001-23, TESLA FEL 2001-05 (2001)} 192p., hep-ph/0106315


\bibitem{Jik}   G.~Jikia, S.~S\"oldner-Rembold,
{\em Nucl. Phys. B (Proc. Suppl.)} {\bf 82} (2000) 373; M.~Melles,
W.J.~Stirling, V.A.~Khoze,{\em Phys.\ Rev.} {\bf D61}(2000)
054015; B.Badelek et al. TESLA Technical Design Report, {\bf p.
VI, chap.1} {\em DESY 2001-011, TESLA Report 2001-23, TESLA FEL
2001-05 (2001)}  p.1-98, hep-ex/0108012


\bibitem{BGI} A.T. Banin, I.F. Ginzburg, I.P. Ivanov, {\em Phys. Rev.}
{\bf D 59} (1999) 115001.

\end{thebibliography}
\end{document}